\documentclass{article}
\usepackage{spconf, amsmath, graphicx}
\usepackage{float}
\usepackage{subfigure}
\usepackage{cite}
\usepackage[urlcolor=blue]{hyperref}

\makeatletter
\newcommand*\bigcdot{\mathpalette\bigcdot@{.5}}
\newcommand*\bigcdot@[2]{\mathbin{\vcenter{\hbox{\scalebox{#2}{$\m@th#1\bullet$}}}}}
\makeatother

\title{towards expressive speaking style modelling with hierarchical context information for mandarin speech synthesis}

\name{Shun Lei$^{1,\dagger}$\thanks{$^{\dagger}$Work conducted when the first author was intern at Huya Inc.}, Yixuan Zhou$^1$, Liyang Chen$^1$, Zhiyong Wu$^{1,3,*}$\thanks{* Corresponding author.},  Shiyin Kang$^2$, Helen Meng$^{1,3}$}
\address{
    $^1$ Shenzhen International Graduate School, Tsinghua University, Shenzhen, China\\
    $^2$ Huya Inc., Guangzhou, China\\
    $^3$ The Chinese University of Hong Kong, Hong Kong SAR, China\\
    \small{
        \{leis21, zhouyx20, cly21\}$@$mails.tsinghua.edu.cn, 
        \{zywu, hmmeng\}$@$se.cuhk.edu.hk,
        \{kangshiyin\}$@$huya.com
    }
}
\begin{document}
\ninept
\maketitle
\begin{abstract}
Previous works on expressive speech synthesis mainly focus on current sentence. 
The context in adjacent sentences is neglected, resulting in inflexible speaking style for the same text, which lacks speech variations.
In this paper, we propose a hierarchical framework to model speaking style from context.
A hierarchical context encoder is proposed to explore a wider range of contextual information considering structural relationship in context, including inter-phrase and inter-sentence relations. 
Moreover, to encourage this encoder to learn style representation better, we introduce a novel training strategy with knowledge distillation, which provides the target for encoder training. 
Both objective and subjective evaluations on a Mandarin lecture dataset demonstrate that the proposed method can significantly improve the naturalness and expressiveness of the synthesized speech\footnote{ Synthesized speech samples are available at: \href{https://thuhcsi.github.io/icassp2022-expressive-tts-hierarchical-context}{https://thuhcsi.github.io/ icassp2022-expressive-tts-hierarchical-context}}.
\end{abstract}


\section{Introduction}
\label{sec:intro}
\begin{figure*}[!htb]
	\centering
	\includegraphics[width=0.80\linewidth]{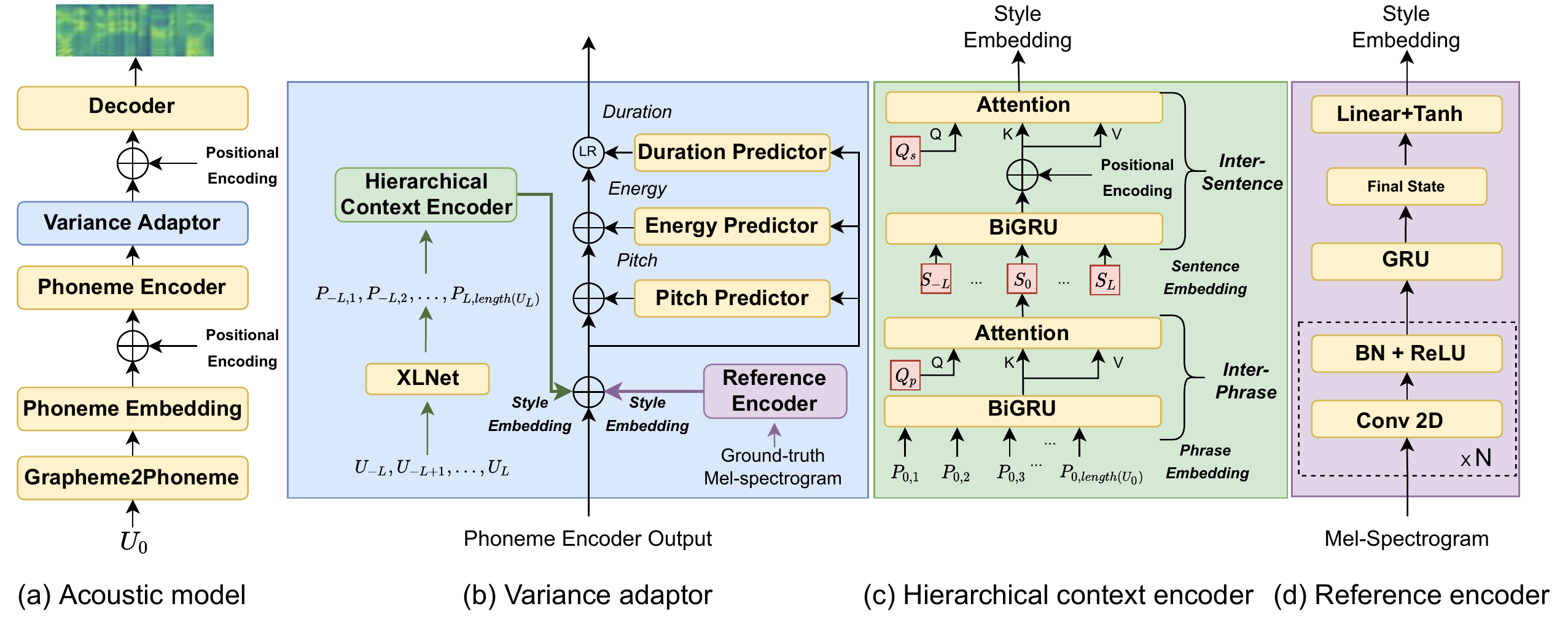}
	\label{fig.model}
	\caption{The overall architecture of the proposed model}
\end{figure*}
Text-to-speech (TTS)
aims to generate intelligible and natural speech from text.
With the development of deep learning, neural network based TTS models can already synthesize high-quality speech with neutral speaking style \cite{tacotron2, deepvoice3, fastspeech2}.
Since speech with repetitive neutral speaking style has gradually fatigued users, speech with richer expressiveness is in great demand.
Therefore, how to model expressive speaking style  
has attracted an increasing interest in academia and industry.

Some recent studies 
devote effort to improve the naturalness and expressiveness of generated speech by modeling speaking style from speech.
In \cite{reference, gst}, a reference encoder is utilized to extract the speaking style representation in an unsupervised manner from a given reference audio.
Similarly, \cite{VAEGST} incorporates the variational autoencoder (VAE) \cite{VAE} into Tacotron 2 \cite{tacotron2} to learn the latent representation with better style disentanglement.
These approaches can generate expressive speech but require auxiliary inputs in inference, such as manually-determined token weights and reference speech.

To deal with this deficiency, the text-predicted global style token (TP-GST) model \cite{TPGST} is proposed to predict the speaking style from text directly.
Benefiting from the great semantic representation ability of the pre-trained language models, such as bidirectional encoder representations from transformer (BERT) \cite{bert}, some latest works \cite{berttacotron, bertemb} use text representation derived from BERT to predict speaking style.
However, it is difficult for TTS model to directly learn speaking style from text in such an implicit way under the situation of limited TTS data.
Meanwhile, the above text-predicted methods only consider the current sentence during synthesis, which is inadequate for modeling expressive speaking style.
For the same input text, these models fail to capture the diverse speech variations (e.g., intonation, rhythm, stress, emotion) that may be brought by the different context of neighbor sentences, resulting in synthesized speech with inflexible speaking style \cite{survey}.
This is against the process of human perception that the speaking style of current speech should be largely influenced by the context.  
Several studies have also demonstrated that considering a wider range of contextual information contributes more to expressive speech synthesis \cite{longformevaluationg, survey}.

In this paper, we propose a hierarchical framework to model the speaking style from context to improve the expressiveness of synthesized speech.
A hierarchical context encoder is introduced to extract more effective context information for predicting the speaking style of the current utterance. 
The encoder tries to explore the structural relationship in context at two levels, one is the inter-phrase relations in a sentence, the other is the inter-sentence relations in a longer context window covering several sentences.
Inspired by knowledge distillation in concurrent work \cite{wsv}, we further introduce a reference encoder to extract style embedding from speech and then use it to explicitly guide the prediction of speaking style rather than in an implicit way.

Both objective and subjective evaluations on a Mandarin lecture dataset demonstrate that the proposed model can synthesize more expressive speech and better model the speech variations.
We also construct a plain context encoder without the use of inter-sentence relations for comparison. 
The evaluation results verify the effectiveness of hierarchical context encoder.

\section{methodology}
\label{sec:method}

The architecture of our proposed model is illustrated in Fig.\ref{fig.model}.
It consists of three major parts: a reference encoder, a hierarchical context encoder and a sequence-to-sequence acoustic model based on FastSpeech 2 \cite{fastspeech2}.
The reference encoder is used to extract the style embedding from reference speech, and the hierarchical context encoder is used to predict the style embedding from context.
Inspired by knowledge distillation, the outputs of the reference encoder are used to guide the training of the hierarchical content encoder.
Then the style embedding is added to the phoneme encoder output and is passed to the variance adaptor of the acoustic model, in order to predict speech variations more accurately and generate speech with expressive speaking style.
The details of each component are as follows.

\subsection{Reference Encoder}
To extract the style embedding from the reference speech, we introduce a reference encoder.
This reference encoder is consistent to the structure proposed in \cite{reference}, composed of a stack of 6 convolutional layers, a gated recurrent unit (GRU) \cite{gru} and a fully-connected layer, as shown in Fig.\ref{fig.model}(d).
All convolutional layers are followed by ReLU activation and batch normalization \cite{ioffe2015batch}, and the fully-connected layer is followed by Tanh activation.
The mel-spectrogram of ground-truth speech corresponding to the given text goes through the reference encoder and the output of the encoder is regarded as the style embedding of this speech.

\subsection{XLNet and Hierarchical Context Encoder}

To improve expressiveness and naturalness of synthesized speech, we combine XLNet and the hierarchical context encoder to predict the speaking style from a fixed number of the \emph{past, current and future sentences}.
Fig.\ref{fig.model}(b) illustrates the 
process of this module.

\subsubsection{XLNet}
XLNet \cite{xlnet} is a pre-trained language model proposed recently, which has achieved superior performance in many natural language processing tasks,
especially in reading comprehension and text classification \cite{nlpsurvey}.
Compared with BERT, XLNet can directly process longer text even at paragraph level without length limitation and thus obtain the semantic information considering a wider range of context.
Hence, we employ XLNet which is more suitable for our task, to derive better text representation as inputs.

Let $L$ be the number of sentences considered in the past or future context.
$U_0$ is defined as the current sentence. $U_{-L}$, $U_{1-L}$, ..., $U_{-1}$ and $U_{1}$, $U_{2}$, ..., $U_{L}$ are the past and future sentences respectively.
All $2L+1$ sentences are firstly concatenated to form a long new text, and then passed to a pre-trained phrase-level XLNet model to obtain phrase embedding sequence, which is described as:
\begin{gather}
    U = Concat(U_{-L},U_{1-L},\cdots,U_{L}), \\
    P_{-L,1},P_{-L,2},\cdots,P_{L,length(U_L)} = XLNet(U),
\end{gather}
where $Concat(\bigcdot)$ is the concatenation operation, $length(U_i)$ is the number of phrases in $U_i$, and $P_{i,j}$ is the 768-dim phrase embedding corresponding to the $j$th phrase of sentence $U_i$ in XLNet outputs.

\subsubsection{Hierarchical Context Encoder}
Phrase-level representations from the pre-trained language model do not explicitly consider structural relationship of context due to the limitation of its training task.
Previous attempts \cite{TPGST, berttacotron} on predicting style in TTS are more about leveraging text in a plain way, which only model the inter-phrase relations by using recurrent neural network (RNN) or attention module.
This is difficult to capture hierarchical long-distance dependencies among phrases and sentences.
To extract more effective context information, we design a hierarchical context encoder for predicting the speaking style of current sentence inspired by \cite{han}.

The hierarchical context encoder contains two levels of attention network, the inter-phrase and the inter-sentence, as shown in Fig.\ref{fig.model}(c).
The two attention networks have similar architecture, which contains a bidirectional GRU \cite{gru} and a scaled dot-product attention module \cite{vaswani2017attention}.
The inter-phrase attention network is used to obtain a sentence-level representation based on each phrase and inter-phrase relations in a sentence.
Firstly, the phrase embeddings in a sentence are passed to a bidirectional GRU to get the phrase-level representations considering temporal relationship.
Not all phrases contribute equally to the meaning of a sentence, so a scaled dot-product attention is then introduced to calculate the weights of each phrase and to aggregate them into a sentence-level embedding. This can be formulated as:
\begin{gather}
    K = H_iW_k, \\
    V = H_iW_v, \\
    S_i = A(Q_p,K,V) = softmax(\frac{Q_pK^T}{\sqrt{d_q}})V,
\end{gather}
where $W_k$ and $W_v$ are linear projection matrices of attention keys and values. $H_i$ is $N\times256$-dim bidirectional GRU outputs of 
phrases in the sentence $U_i$,
and $N=length(U_i)$.
$Q_p$ is a 256-dim vector, which can be seen as a high level representation of a fixed query 
``to what extent does the phrase influence the speaking style".
$Q_p$ is randomly initialized and learnable during the training.
$S_i$ is the 256-dim sentence embedding of sentence $U_i$.

Similarly, the inter-sentence attention network is used to predict a speaking style embedding 
based on sentence embeddings and inter-sentence relations in context.
Compared with inter-phrase attention network, an additional positional encoding \cite{transformer},
which is found to have a great influence on performance in our trials, 
is added to the corresponding output of bidirectional GRU to provide the relative position information of sentences.
Finally, the inter-sentence attention network outputs the speaking style embedding of current utterance. It is noteworthy that this hierarchical structure has encoded inter-phrase and inter-sentence relations, which contributes to be a hierarchical information aware model.

\subsection{Acoustic Model}
We adopt FastSpeech 2 as the acoustic model and make several changes as illustrated in Fig.\ref{fig.model}.
Firstly, the style embedding from the reference encoder or hierarchical context encoder is replicated to phoneme level, and is added to the outputs of phoneme encoder, and then is passed to the variance predictors as shown in Fig.\ref{fig.model}(b).
This modification allows the variance predictors to predict duration, pitch and energy more accurately.
Secondly, the length regulator is moved after the variance predictors, 
in order to predict variations 
at phoneme level rather than frame level, which proves to further improve speech quality \cite{fastpitch}.

\subsection{Model Training}
Generally, it is challengeable for acoustic model to implicitly learn speaking style from text under the situation of limited TTS data.   
To encourage the hierarchical context encoder to learn style representation better, the proposed model is trained with the knowledge distillation strategy in three steps.

In the first step, the acoustic model and the reference encoder are jointly trained with paired $<$utterance, speech$>$ data to get a well-trained speech style extractor in an unsupervised way.
Then the style embeddings extracted from all speeches in the training set can be regarded as ground-truth speaking style representation. 
In the second step, we leverage knowledge distillation to transfer the knowledge from the reference encoder to the hierarchical context encoder.
That is, we use ground-truth style embeddings as target to guide the prediction of speaking style representation from context, for training the hierarchical context encoder only.
Finally, we jointly train the acoustic model and hierarchical context encoder with a lower learning rate to further improve the naturalness of synthesized speech.

\section{Experiments}
\label{sec:exp}
\subsection{Training Setup}
All the models are trained on an internal single-speaker Mandarin lecture dataset, which contains around 7 hours lecture speech spoken by a male Mandarin native speaker.
The speaking styles vary among utterances, and the speed, pitch and energy fluctuate greatly in an utterance.
The dataset has a total of 4700 audio clips, of which 200 clips are used for validation and 100 clips for test, and the rest is for training.

For feature extraction, we transform the raw waveforms into 80-dim mel-spectrograms with sampling rate 24kHz, frame size 1200 and hop size 240. 
Silences at the beginning and the end of each utterance are trimmed.
The phoneme duration are extracted by Montreal Forced Aligner \cite{mfa} tool.
An open-source pre-trained Chinese phrase-level XLNet-base model\footnote[1]{\href{https://github.com/ymcui/Chinese-XLNet}{https://github.com/ymcui/Chinese-XLNet}}
is used in our experiments.
The context of current sentence is made up of its two past sentences, two future ones and itself.
And the raw text is converted to phoneme sequence with some prosody tags as the input of acoustic model.

We take 180k steps to train the acoustic model and reference encoder, and then take 20k steps to train the hierarchical context encoder and 20k steps to adapt the hierarchical context encoder and acoustic model.
All the trainings are conducted with a batch size of 16 on a NVIDIA V100 GPU.
The Adam optimizer is adopted with $\beta_1=0.9$ , $\beta_2=0.98$.
The warm-up strategy is used before 4000 steps.
In addition, a well-trained HIFI-GAN\cite{kong2020hifi} is used as the vocoder to generate waveform.

\subsection{Compared Methods}
Two FastSpeech 2 based models are implemented for comparison, and the details are described as follows: \\
\textbf{FastSpeech 2:}
Original FastSpeech 2 \cite{fastspeech2} with several changes consistent to the proposed model as described in section 2.3.  
\\
\textbf{XLNet-FastSpeech 2:}
Inspired by \cite{berttacotron}, we set an end-to-end TTS model by combining XLNet with FastSpeech 2, which also considers context information.
In XLNet-FastSpeech 2, we construct a plain context encoder without the use of inter-sentence relations.
The same phrase embeddings obtained from XLNet are directly passed to a GRU layer whose final state is used as a style embedding.
In this way, the context information is considered in a plain manner rather than hierarchical structure.

\subsection{Subjective Evaluation}

We conduct the mean opinion score (MOS) test to evaluate the naturalness and expressiveness of generated speeches.
The sentences are selected randomly in test set.
20 native Chinese speakers are asked to listen generated speeches and rate on a scale from 1 to 5 with 1 point interval.
The MOS results are shown in Table \ref{tab:mos}.
It can be observed that there exists a large gap between FastSpeech 2 and Ground Truth, indicating that training TTS model with an expressive dataset is difficult.
Our proposed approach achieves the best MOS of $4.067$, exceeding FastSpeech 2 by $0.388$ and XLNet-FastSpeech 2 by $0.225$.
Some subjects also report that the speeches synthesized by the proposed approach have richer expressiveness, especially in intonation, rhythm and stress. 

\begin{table}[th]
\renewcommand{\arraystretch}{1.0}
  \caption{The naturalness and expressiveness MOS of different models with 95\% confidence intervals.}
  \label{tab:mos}
  \centering
  \begin{tabular}{l|c} 
    \toprule
    \textbf{Model} &\textbf{MOS} \\
    \midrule
    Ground Truth & $4.754\pm0.066$ ~~~               \\
    FastSpeech 2 & $3.679\pm0.099$ ~~~  \\
    XLNet-FastSpeech 2 & $3.842\pm0.096$ ~~~  \\
    Proposed & $\mathbf{4.067\pm0.089}$ ~~~ \\
    \bottomrule
  \end{tabular}
\end{table}

\begin{figure}[!htb]
	\centering
	\includegraphics[width=0.8\linewidth, height=0.25\linewidth]{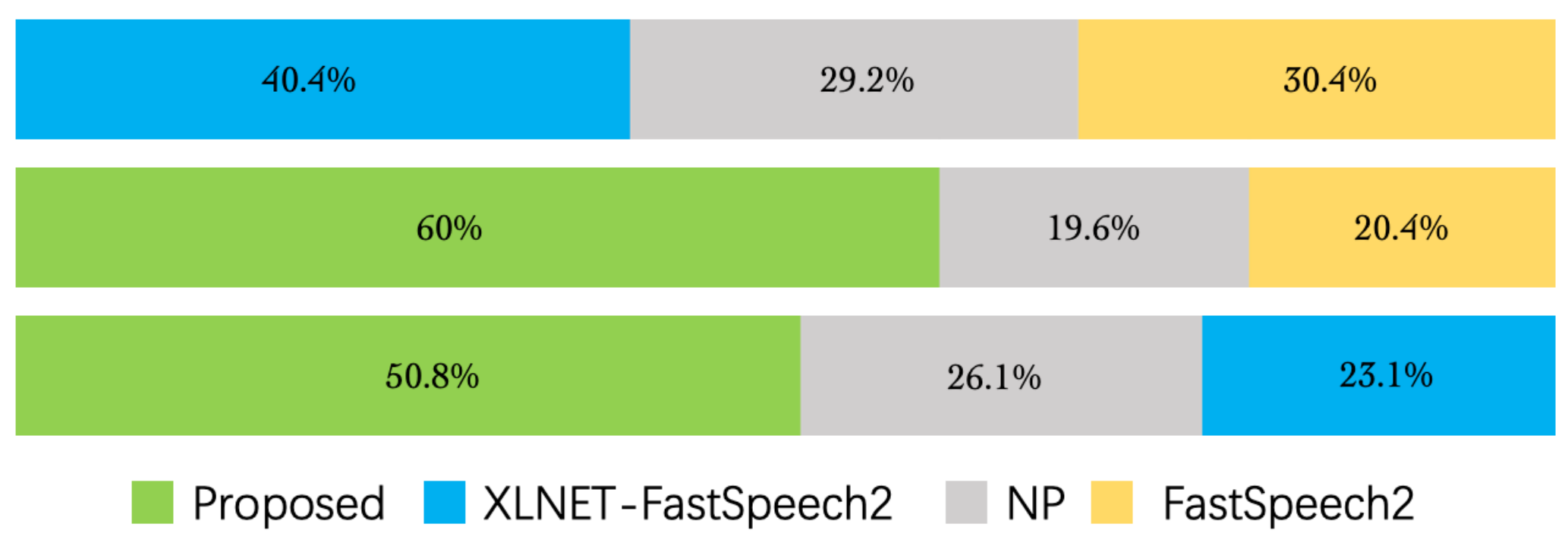}
	\caption{Result of the ABX test for naturalness and expressiveness. NP means no preference.}
	\label{fig:abx}
\end{figure}

ABX preference test is also conducted on our proposed model and two baselines.
The same 20 subjects are asked to choose a preferred speech in terms of naturalness and expressiveness 
between a pair of methods.
As shown in Fig.\ref{fig:abx},
the preference rate of our proposed model exceeds FastSpeech 2 by $39.6\%$ and XLNet-FastSpeech 2 by $27.7\%$ respectively, and XLNet-FastSpeech 2 is preferred than FastSpeech 2.

Both MOS and ABX preference tests demonstrate that our proposed method significantly outperforms the two baselines in terms of naturalness and expressiveness.
Compared with FastSpeech 2 that only uses phoneme sequence with prosody tags as input, both our proposed model and XLNet-FastSpeech 2 perform better, indicating that considering contextual information is really helpful for expressive speech synthesis. 
Our proposed model also makes further improvement compared with XLNet-FastSpeech 2 that simply uses contextual information in a plain manner.
It demonstrates that introducing the hierarchical context encoder and leveraging knowledge distillation strategy can model speaking style for TTS better.

\subsection{Objective Evaluation}
We employ the root mean square error (RMSE) of pitch and energy, and the mean square error (MSE) of duration and mel cepstral distortion (MCD) as the objective evaluation metrics 
followed by previous works \cite{fastspeech2}.
To calculate F0 and energy RMSE, we first use the dynamic time warping (DTW) to construct the alignment paths between the predicted mel-spectrogram and the ground-truth one.
Then, the F0 sequence and energy sequence are aligned towards ground-truth following the DTW path.
We also utilize DTW to compute the minimum MCD by aligning the two sequences.
For duration, we compute the MSE between the predicted duration and ground-truth duration.

\begin{table}[th]\footnotesize
\renewcommand{\arraystretch}{1.0}
  \caption{Objective evaluations for different models.}
  \label{tab:objective}
  \centering
  \begin{tabular}{lccc} 
    \toprule
    \textbf{} & \textbf{FastSpeech 2} & \textbf{XLNet-FastSpeech 2} & \textbf{Proposed} \\
    \midrule
    F0 RMSE  & $54.395$ & $53.921$ & \textbf{51.871} ~~~  \\
    Energy RMSE  & $3.733$ & $3.348$ & \textbf{3.138} ~~~  \\
    Duration MSE & $0.1267$ & $0.1237$ & \textbf{0.1175} ~~~ \\
    MCD  & $4.614$ & $4.601$ & \textbf{4.535} ~~~ \\
    \bottomrule
  \end{tabular}
\end{table}

The evaluation results of different models on the test set are shown in Table \ref{tab:objective}.
It is observed that our proposed model outperforms the two baselines in all objective evaluation metrics.
The results indicates that our proposed model can synthesize speech closer to the ground-truth, especially in the speech variations, such as pitch, energy and duration.

\subsection{Ablation Study}
We conduct two ablation studies to demonstrate the effectiveness of knowledge distillation training strategy and hierarchical context encoder respectively.
Comparison mean opinion score (CMOS)
is employed to compare the synthesized speeches in terms of naturalness and expressiveness.

When removing the knowledge distillation strategy from the proposed method, that is, predicting the style embedding without the guidance of reference encoder, we find it results in -0.346 CMOS. 
This indicates that learning speaking style representation from context in an explicit way is more suitable for this task. 

In addition,
we further remove the inter-sentence part in hierarchical context encoder, that is, predicting the speaking style by a plain context encoder, we find it results in -0.609 CMOS, providing the importance of considering structural relationship in context for expressive speech synthesis.

\subsection{Case Study}

To explore the impact of contextual information on the expressiveness of synthesized speech, a case study is conducted to synthesize the same utterance with different context: i) using ground-truth context (original context); ii) randomly selecting 4 sentences and itself as context (irrelevant context); iii) using current sentence only (no context).

\begin{figure}[!htb]
	\centering
	\includegraphics[width=0.8\linewidth, height=0.24\linewidth]{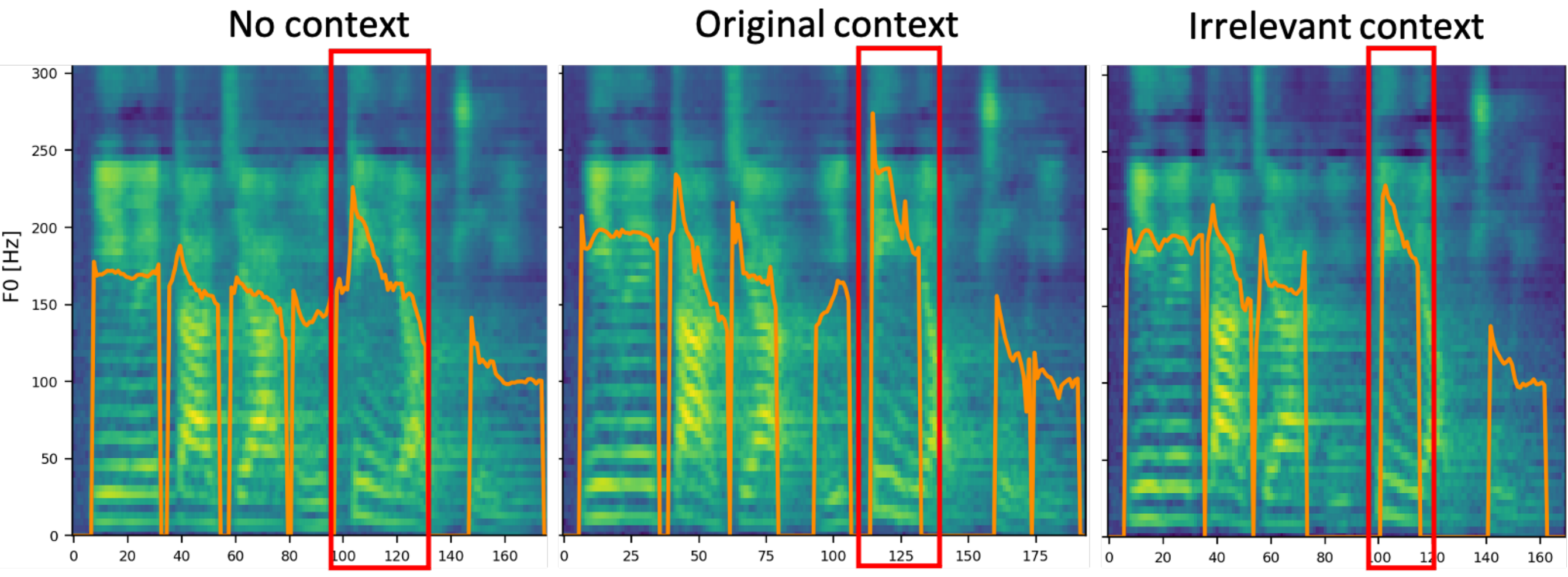}
	\caption{The mel-spectrograms and pitch contours of speech synthesized by proposed model with different context. The text means ``We must care" and the red box corresponds to ``must" in Chinese.}
	\label{fig:casestudy}
\end{figure}

As shown in Fig.\ref{fig:casestudy}, the mel-spectrograms, pitch contours and duration of these speeches are obviously different.
The speech generated with original context contains richer pitch variations overall than others.
Especially, it can be found that speech generated with original context emphasizes on the word ``must" (the red box) through a higher pitch value, which is more in line with ground truth.
Compared original text with no context, it shows that considering contextual information is necessary to model speaking style.
When comparing original text with irrelevant context, it shows that valid contextual information is helpful rather than arbitrary semantic information.
Our model successfully learns the meaningful contextual information in modeling speaking style, providing the effectiveness of the proposed method.
The result of the case study demonstrates that modeling speaking style from context can effectively affect the stress, pitch and duration of synthesized speech.
Thus, the same utterance can be synthesized in a variety of ways according to context, to achieve more natural and expressive speech synthesis.

\section{Conclusions}
\label{sec:conclusions}

In this paper, we propose a hierarchical framework to model the speaking style from context for expressive speech synthesis.
A hierarchical context encoder is introduced to better utilize contextual information.
A novel training strategy with knowledge distillation is used to further improve the performance of style prediction in TTS.
Experimental results in both objective and subjective evaluations demonstrate that our proposed approach could significantly improve the naturalness and expressiveness of the synthesized speech.  
In ablation studies, the effectiveness of hierarchical context encoder and knowledge distillation has also been proved.

\textbf{Acknowledgement}: This work was supported by National Key R\&D Program of China (2020AAA0104500), National Natural Science Foundation of China (NSFC) (62076144) and National Social Science Foundation of China (NSSF) (13\&ZD189).

\vfill\pagebreak
\ninept

\bibliographystyle{IEEEbib}
\bibliography{strings,refs}

\end{document}